\documentclass{revtex4}
\begin{document}
\title{Pathological behaviour of the scalar graviton in Ho\v{r}ava-Lifshitz gravity}
\vfill
\author{Kazuya Koyama\footnote{Kazuya.Koyama@port.ac.uk}}
\author{Frederico Arroja\footnote{arrojaf@yukawa.kyoto-u.ac.jp}}

\affiliation{
$*$Institute of Cosmology and Gravitation, University of Portsmouth, Portsmouth PO1 3FX, UK. \\
$\dagger$Yukawa Institute for Theoretical Physics, Kyoto University, Kyoto 606-8502, Japan.
}

\vfill
\date{today}

\begin{abstract}
We confirm the recent claims that, in the infrared limit of Ho\v{r}ava-Lifshitz gravity, the scalar graviton becomes a ghost if the sound speed squared is positive on the flat de Sitter and Minkowski background. In order to avoid the ghost and tame the instability, the sound speed squared should be negative and very small, which means that the flow parameter $\lambda$ should be very close to its General Relativity (GR) value. We calculate the cubic interactions for the scalar graviton which are shown to have a similar structure with those of the curvature perturbation in k-inflation models. The higher order interactions become increasing important for a smaller sound speed squared, that is, when the theory approaches GR. This invalidates any linearized analysis and any predictability is lost in this limit as quantum corrections are not controllable. This pathological behaviour of the scalar graviton casts doubt on the validity of the projectable version of the theory.
\end{abstract}
\maketitle

\section{Introduction}
There is a growing interest in a theory proposed by Ho\v{r}ava as an ultraviolet (UV) renormalizable gravity theory \cite{Horava:2009uw}. This is inspired by the Lifshitz theory in solid state physics
and is often called Ho\v{r}ava-Lifshitz gravity. The essential ingredient of the theory is the breaking of Lorentz invariance. This is achieved by assuming the existence of a preferred foliation by 3-dimensional time-constant hypersurfaces which separates space and time. Then the action that contains higher spatial derivatives of the metric is introduced, which improves the UV behaviour of the graviton propagator. In this theory, general covariance is broken but it is still invariant under the foliation-preserving diffeomorphism, $x^i \to \tilde{x}^i(t, x^i), t \to \tilde{t}(t)$. We consider the Arnowitt-Deser-Misner decomposition of space-time \cite{Arnowitt:1960es}
\begin{equation}
ds^2 = - (N^2 -N_i N^i) dt^2 + 2 N_i dx^i dt + \gamma_{ij} dx^i dx^j.
\end{equation}
Here the lapse $N$ represents the gauge degree of freedom associated with the space independent time reparametrisation $t \to \tilde{t}(t)$. Thus the lapse function $N$ is
restricted to be a function of time $N=N(t)$. This is called the projectability condition and this is the key property of the theory. We should note that the theory without the projectability condition has also been studied but it was shown to have theoretical inconsistences \cite{Charmousis:2009tc, Blas:2009yd,Li:2009bg}.
The action is given by
\begin{equation}
S = \frac{M_{\rm pl}^2}{2}
\int dt d^3 x N \sqrt{-\gamma} \left(
\left(K^{ij} K_{ij} - \lambda K^2 \right) + {}^{(3)} R - 2 \Lambda + {\cal L}_V \right),
\end{equation}
where $\lambda$ is the flow parameter which is supposed to become one in the infrared (IR) limit, $ {}^{(3)}R$ is
the three dimensional Ricci scalar, $\Lambda$ is the cosmological constant and the
extrinsic curvature is defined as
\begin{equation}
K_{ij} = \frac{1}{2 N} (\dot{\gamma}_{ij} - \nabla_i N_j - \nabla_j N_i),
\end{equation}
where a dot indicates a derivative with respect to time and $K=\gamma^{ij}K_{ij}$. From now on, we set the Planck scale $M_{\rm pl}=1$.
The potential term ${\cal L}_V$ contains higher-order spatial derivatives which are important in the UV limit to ensure the power-counting renormalizability of the theory but it can be neglected in the IR limit.

In the IR limit, the theory is supposed to flow into General Relativity (GR), $\lambda \to 1$.
However, due to the projectability condition, there is a crucial difference between this theory and GR. Since the general covariance is broken, a new scalar degree of freedom in the graviton inevitable arises \cite{Horava:2008ih,Horava:2009uw,Kobakhidze:2009zr}. Recently, it has been pointed out that this scalar mode shows a pathological behaviour \cite{Horava:2008ih,Horava:2009uw, Blas:2009yd, Blas:2009qj}.
The aim of this paper is to clarify the issue of the scalar graviton in the projectable version of the Ho\v{r}ava-Lifshitz gravity on the Minkowski and de Sitter backgrounds.

This paper is organized as follows. In section II, we use the Hamiltonian formulation
and identify the scalar graviton in the Minkowski background spacetime. We confirm the finding of
Ho\v{r}ava \cite{Horava:2008ih, Horava:2009uw} (see also Blas \emph{et al.} \cite{Blas:2009yd} and Sotiriou \emph{et al.} \cite{Sotiriou:2009bx}) that the scalar graviton is a ghost if the sound speed squared is positive.
In order to make the scalar graviton healthy, the sound speed squared must be negative
and it is inevitably unstable. Thus the sound speed squared should be close to zero, which
means $\lambda \to 1$. In section III, we derive the cubic order interactions for the scalar graviton using the approach pioneered by Maldacena \cite{Maldacena:2002vr}. We show that in the small sound speed limit, which is necessary to avoid the instability, the cubic interactions
are important at very low energies. This invalidates any linearized analysis and
any predictability is lost due to unsuppressed loop corrections.
Section IV is devoted to discussions.

\section{Scalar graviton in the Hamiltonian formulation}
In order to identify the scalar graviton in the IR limit of the Ho\v{r}ava-Lifshitz gravity, it is useful to use the Hamiltonian formulation. In this section, we consider perturbations around the Minkowski spacetime to make the derivation as simple as possible to capture the basic picture. We should mention that this analysis was done already in
Ref.~\cite{Blas:2009qj} (see also \cite{Cai:2009dx}) but it is instructive to repeat their analysis using the Hamiltonian formulation to understand the origin of the scalar graviton. We perturbed the metric as
\begin{equation}
N= 1+ \alpha(t), \quad N_i =  \partial_i \beta, \quad
\gamma_{ij} = \delta_{ij} + 2 \left(
\delta_{ij}+k^{-2} \partial_i \partial_j \right) \zeta
- 2 k^{-2}  \partial_i \partial_j \chi,
\end{equation}
where $k^{-2}$ is the inverse of $k^2\equiv-\partial^2$.
Then the second order Lagrangian is obtained as \cite{Blas:2009qj}
\begin{equation}
{\cal L} =\frac{1}{2}
\left[
-2 \dot{\zeta}^2 + 2 k^2 \zeta^2 + 4 \alpha k^2 \zeta
 -4 k^2 \dot{\zeta} \beta - 4 \dot{\zeta} \dot{\chi} -
(\lambda-1) \left(k^2 \beta + \dot{\chi} + 2 \dot{\zeta} \right)^2
\right].
\end{equation}

In order to extract the dynamical degrees of freedom, it is transparent to use the Hamiltonian formulation. We closely follow the approach developed in Ref.~\cite{Garriga:1997wz}.
The conjugate momenta are obtained as
\begin{eqnarray}
\pi_{\alpha} &=& 0, \quad \pi_{\beta}=0, \\
\pi_{\zeta} &=& - 2 (\dot{\zeta}+\dot{\chi}+ k^2 \beta) - 2(\lambda-1) (k^2 \beta +\dot{\chi} + 2 \dot{\zeta}),  \\
\pi_{\chi} &=& - 2 \dot{\zeta} - (\lambda-1) (k^2 \beta +\dot{\chi} +  2 \dot{\zeta}).
\end{eqnarray}
Then the Lagrangian can be rewritten as
\begin{equation}
{\cal L} = \pi_{\zeta} \dot{\zeta} + \pi_{\chi} \dot{\chi} - {\cal H}
- \beta C_{\beta} - \alpha(t) C_{\alpha},
\end{equation}
where
\begin{equation}
 C_{\alpha} = - 2 k^2 \zeta, \quad C_{\beta} =  - k^2 \pi_{\chi},
\end{equation}
and
\begin{equation}
{\cal H} = -k^2 \zeta^2 + \frac{1}{4 ( 3 \lambda-1)}
\left[ 2 (2 \lambda-1) \pi_{\chi}^2 - 4 \lambda \pi_{\chi} \pi_{\zeta} + (\lambda-1) \pi_{\zeta}^2 \right].
\end{equation}
In GR, $C_{\alpha}$ and $C_{\beta}$ are both constraints and they imply $
k^2 \zeta=0$ and $\pi_{\chi}=0$, then ${\cal H}=0$ as $\lambda=1$. Thus there is no dynamical degree of freedom left in the scalar sector. Of course, this is because there is no scalar graviton in GR thanks to general covariance. However,
unlike in GR, $C_{\alpha}$ is not a constraint in this theory. Since $\alpha(t)$ does not depend on spatial coordinates, the action that is proportional to $\alpha(t)$ is given by
\begin{equation}
-2 \alpha(t) \int d^3 x \partial^2 \zeta.
\end{equation}
This reduces to a surface term and we get no constraint. On the other hand $C_{\beta}=0$ still
holds and this comes from the residual 3-dimensional diffeomorphisms. Using this constraint, the Lagrangian is reduced to
\begin{equation}
{\cal L} = \pi_{\zeta} \dot{\zeta} - {\cal H}, \quad
{\cal H} = -\frac{c_{\zeta}^2}{4 } \pi_{\zeta}^2 - k^2 \zeta^2,
\label{Hami}
\end{equation}
where
\begin{equation}
c_{\zeta}^2 = \frac{1-\lambda}{3\lambda-1},
\end{equation}
is the sound speed squared.
We should note that the $\lambda=1$ case looks already pathological as $\pi_{\zeta}$ does not appear in the quadratic Hamiltonian. This result agrees with the one obtained by the St\"{u}ckelberg approach where it is argued that the lack of $\pi_{\zeta}$ in the quadratic Hamiltonian would indicate that quantum fluctuations of $\pi_{\zeta}$ are unsuppressed and any interactions containing $\pi_{\zeta}$ would blow up \cite{Blas:2009yd}.
A pathology with $\lambda=1$ was also found in a different context in Ref.~\cite{Mukohyama:2009tp}.
In the following we assume $\lambda \neq 1$. We discuss the $\lambda=1$ case in Appendix \ref{App:lambdaone}.

Using the Hamilton's equation
\begin{equation}
\dot{\zeta} = -\frac{c_{\zeta}^2}{2 } \pi_{\zeta},
\label{dzeta}
\end{equation}
we can write down the Lagrangian in terms of $\dot{\zeta}$ as
\begin{equation}
{\cal L} = -\frac{1}{c_{\zeta}^2} \dot{\zeta}^2 + k^2 \zeta^2.
\end{equation}
Then the equation of motion for the scalar graviton is obtained as
\begin{equation}
\ddot{\zeta} + c_{\zeta}^2 k^2 \zeta =0.
\end{equation}
This result agrees with Ref.~\cite{Wang:2009yz}.

Then the second order action can be written as
\begin{eqnarray}
S_2 &=& -\int dt d^3 x \left[\frac{1}{c_{\zeta}^2} \dot{\zeta}^2 - (\partial \zeta)^2 \right].
\end{eqnarray}
We immediately notice that the action has a wrong over-all sign. In order not to have an instability, $c_{\zeta}^2>0$, then the coefficient in front of the time kinetic term becomes negative and the scalar mode becomes a ghost. This is clearly seen from the Hamiltonian (\ref{Hami}). For $1/3< \lambda<1$, $c_{\zeta}^2>0$ but then the Hamiltonian is negative definite \cite{Sotiriou:2009bx,Bogdanos:2009uj}. Note that the sign of the time derivative term agrees with Ho\v{r}ava's original calculation (see Eq. (4.56) of \cite{Horava:2008ih}). When $\zeta$ is not a ghost, then it is unstable because $c_{\zeta}^2 <0$. In this case, we need to take into account the higher derivative terms to address the fate of the instability, but this means that the time scale of the instability is at least $1/|c_{\zeta}| M$, where $M$ is the UV scale where the higher derivative terms become important. In order not to have the instability within the age of Universe for example, one needs $|c_{\zeta}| \sim H_0/M$. Thus we should make $\lambda$ very close to one or the UV scale of the theory very low. This is the finding of Ref.~\cite{Blas:2009qj}.
It is known that if the sound speed squared is small, the higher order interactions become increasingly important.
This was also pointed out by Ref.~\cite{Blas:2009qj}, but in the next section, we explicitly calculate the cubic interactions of the scalar graviton to confirm this claim.

\section{Cubic interactions}
Although the Hamiltonian formulation gives a transparent way of identifying the scalar graviton, there is an easier way to derive the higher order interactions. It is the approach pioneered by Maldacena \cite{Maldacena:2002vr},
which is now a common tool to derive higher order actions for inflaton perturbations.
In this section, we consider the de Sitter background by introducing a cosmological constant but we do not include matter. The dynamical equation for the scale factor is given by
\begin{equation}
(3 \lambda -1) \left(\dot{H} + \frac{3}{2} H^2\right) = \Lambda.
\end{equation}
This admits the solution
 \begin{equation}
 \frac{1}{2} \left(
 3 \lambda -1
 \right)
H^2 = \frac{8 \pi G}{3} \rho_0 a^{-3} + \frac{\Lambda}{3},
\label{Friedman}
\end{equation}
where $\rho_0$ is an integration constant . In GR, the Hamiltonian constraint implies $\rho_0=0$, but in the projectable version of the theory, the Hamiltonian constraint is replaced by a global constraint
\begin{equation}
\int d^3 x a^3 \left(\frac{1}{2} (3 \lambda -1) H^2 -\frac{\Lambda}{3}
\right)  =0.
\end{equation}
Thus we do not necessarily have to choose $\rho_0=0$ in our Universe \cite{Mukohyama:2009mz}.
This is an interesting possibility but in the following, we set $\rho_0=0$ and consider the
de Sitter and Minkowski background spacetimes.

We again start from the general action:
\begin{equation}
S= \frac{1}{2} \int dt d^3x \sqrt{\gamma} N(t)\left[(K_{ij} K^{ij} - \lambda K^2)
+{}^{(3)} R - 2 \Lambda \right],
\end{equation}
where we neglected the higher-derivative terms. The momentum constraint is given by
\begin{equation}
\nabla_j K_{i}^{j} - \lambda \nabla_i  K=0.
\end{equation}
By using the three dimensional diffeomorphism and the time reparametrisation invariance, we are allowed to use the following gauge
\begin{equation}
h_{ij} = a(t) ^2e^{2 \zeta} \delta_{ij}, \quad N_i =\partial_i \beta, \quad N(t)= 1.
\end{equation}

At the leading order, the momentum constraint is solved as
\begin{equation}
\partial^2 \beta  =
- a^2 \frac{1}{c_{\zeta}^2} \dot{\zeta}.
\label{betasol}
\end{equation}
Note that if $\lambda=1$, the momentum constraint implies $\dot{\zeta}=0$ and we cannot determine $\beta$. This agrees with the result obtained in the Hamiltonian formalism
Eq.~(\ref{dzeta}). We treat the $\lambda=1$ case separately in Appendix \ref{App:lambdaone} and we assume
$\lambda  \neq 1$ in the following analysis.
Using an expression for the three dimensional Ricci scalar curvature,
\begin{equation}
{}^{(3)}R = -2 a^{-2} e^{-2 \zeta} (\partial_i \zeta \partial^i \zeta+2 \partial^2 \zeta),
\end{equation}
we can easily obtain the second order action for $\zeta$ as
\begin{eqnarray}
S_2 &=& \int dt d^3 x \left[-a^3 \frac{1}{c_{\zeta}^2} \dot{\zeta}^2 + a (\partial \zeta)^2 \right].
\label{S2}
\end{eqnarray}
In the same way, we can calculate the third order action
\begin{equation}
S_3= \int dt d^3 x
\left[
a \zeta \partial_i \zeta \partial^i \zeta - \frac{3 a^3}{c_{\zeta}^2} \zeta \dot{\zeta}^2
+\frac{3}{2a} \zeta \Big(\partial_i \partial_j \beta\partial^i \partial^j \beta -(\partial^2 \beta)^2 \Big) - \frac{2}{a} \partial^2 \beta\partial_i \zeta\partial^i \beta
 \right].\label{S3}
\end{equation}
In Appendix \ref{App:Kinflation}, we compare these results with those in k-inflation models and it is found that these
actions have remarkable similarities to those for the curvature perturbation in k-inflation models with
arbitrary sound speed. In k-inflation models, it is known that higher-order interactions are increasingly
important in a small sound speed limit \cite{Seery:2005wm, Chen:2006nt}.
Thus we expect that we have the same problem in this model.

Now we are in a position to discuss the strong coupling problem. By restoring the Planck scale and using the canonically normalized variable $\zeta_c = M_{\rm pl}c_\zeta^{-1/2}\zeta$, we find that the cubic interactions written by $\beta$ are suppressed only by $M_{\rm pl}c_\zeta^{3/2}$ compared with the quadratic action. Then in the $c_{\zeta} \to 0$ limit, these cubic interactions blow up. In fact, in this limit the quantum gravity scale would be effectively zero. This agrees with the finding in Ref.~\cite{Blas:2009yd}. The origin of this pathology is the projectable condition. In the $\lambda \to 1$ limit, the momentum constraint gives $\dot{\zeta}=0$. In GR, there is also a Hamiltonian constraint and this implies $\partial^2 \zeta=0$. Then $\zeta$ is not a dynamical variable. In this case, it is easy to see that the cubic interactions written by $\beta$ disappear as one can perform an integration by parts. The crucial consequence of the projectability condition is to abandon the (local) Hamiltonian constraint. Then $\zeta$ becomes a dynamical degree of freedom, the cubic interactions given by $\beta$ do not disappear and these interactions blow up in the $c_{\zeta} \to 0$ limit.

In order to see this fact clearly, we apply again the Hamiltonian formulation to
Eq.~({\ref{S2}}). The conjugate momentum is obtained as
\begin{equation}
\pi_{\zeta}= -\frac{2 a^3 \dot{\zeta}}{c_{\zeta}^2}.
\end{equation}
We find that the solution for $\beta$ from the moment constraint, Eq.~(\ref{betasol}), is written as
\begin{equation}
\partial^2 \beta = \frac{\pi_{\zeta}}{2 a}.
\end{equation}
Using the conjugate momentum, the actions can be written as
\begin{eqnarray}
S_2 &=& \int dt d^3 x \left[\dot{\zeta} \pi_{\zeta} -
\left(-\frac{c_{\zeta}^2}{4a^3} \pi_{\zeta}^2 - a (\partial \zeta)^2 \right)
 \right],
 \label{S2ham}
\end{eqnarray}
\begin{equation}
S_3= \int dt d^3 x
\left[
a \zeta \partial_i \zeta \partial^i \zeta - \frac{3c_{\zeta}^2}{4a^3} \zeta \pi_{\zeta}^2
+\frac{3}{2a} \zeta \Big(\partial_i \partial_j \beta\partial^i \partial^j \beta -(\partial^2 \beta)^2 \Big) - \frac{2}{a} \partial^2 \beta\partial_i \zeta\partial^i \beta
 \right].
\label{S3ham}
\end{equation}
In Appendix \ref{App:lambdaone}, we show that these formulae can be applied to both $c_{\zeta}=0$ and $c_{\zeta} \neq 0$ cases. Then we can see the problem from the fact that the momentum $\pi_{\zeta}$ disappears in the $c_{\zeta} \to 0$ limit in the quadratic Hamiltonian but the cubic interactions contain $\pi_{\zeta}$.
This means that the quantum fluctuations of $\pi_{\zeta}$ are unsuppressed and the cubic interactions blow up. This result agrees with the finding in Ref.~\cite{Blas:2009yd} using the St\"uckelberg approach though a precise comparison needs to be done.

\section{Conclusions}
In this paper, we derived the quadratic and cubic order actions for the scalar graviton in the IR limit of the projectable version of the Ho\v{r}ava-Lifshitz gravity. Our findings are fully consistent with
Ref.~\cite{Blas:2009yd, Blas:2009qj}. Firstly,
the scalar graviton becomes a ghost if the sound speed squared $c_{\zeta}^2$ is positive on the flat de Sitter and Minkowski background. In order to avoid the ghost and tame the instability, the sound speed squared should be negative and very small, which means that the flow parameter $\lambda$ should be very close to its GR value. Secondly, in the $c_{\zeta}^2 \to 0$ limit, the cubic interaction terms become increasingly important. In terms of the canonically normalized variable, the cubic interaction terms are suppressed only by $M_{\rm pl}c_\zeta^{3/2}$. Thus the quantum gravity scale becomes effectively zero in the $c_{\zeta} \to 0$ limit. This invalidates any linearized analysis and any predictability is lost in this limit as quantum corrections are uncontrollable. Using the Hamiltonian formulation and writing down the actions in terms of the dynamical variable $\zeta$ and its conjugate momentum $\pi_{\zeta}$, the actions for the scalar graviton can be written in an unified way for $c_{\zeta}=0$ and $c_{\zeta} \neq 0$ cases. The origin of the problem is that the momentum $\pi_{\zeta}$ disappears in the quadratic Hamiltonian in the $c_{\zeta} \to 0$ limit and the quantum fluctuations of $\pi_{\zeta}$ are unsuppressed. The cubic interactions contain $\pi_{\zeta}$ and then they blow up in this limit. In principle, we can continue our analysis and study higher-order interactions. At the fourth order, the scalar perturbations couple to vector and tensor perturbations and the calculations become very complicated (see Ref.~\cite{Arroja:2008ga} for the k-inflation case). We leave this analysis for future work.

Our results cast doubt on the projectable version of Ho\v{r}ava-Lifshitz gravity because one loses predictability
when the theory is supposed to approach GR in the IR limit due to uncontrollable quantum corrections. We should emphasize that these problems have nothing to do with the potential terms which are crucial to achieve the renormalizability of the theory in the UV limit.
General forms of this potential term was proposed in Ref.~\cite{Sotiriou:2009gy} and their linear
perturbations in terms of
$\zeta$ were calculated in Ref.~\cite{Wang:2009yz}. These potential terms are given by the three dimensional Ricci tensor and thus they contain only spatial derivatives of $\zeta$.
The problem found in Refs.~\cite{Blas:2009yd, Blas:2009qj}
and in this paper comes from the kinetic terms that contain time derivatives of $\zeta$. Thus the potential terms cannot cure the problem. It is interesting to find a cure to this problem along the lines of the recent proposal in \cite{Blas:2009qj} in the non-projectable version of the theory.
Note that we did not consider the coupling to matter in our analysis (see for example Refs.~{\cite{Kobayashi:2009hh, Wang:2009az} for cosmological perturbation analysis in the projectable version of the theory). It is straightforward to include an additional scalar field in our analysis and it would be interesting to study its consequences once the pathology of the scalar graviton is cured.

{\it Note added:} While we were preparing this paper, Ref.~\cite{Chen:2009vu} appeared in the arXiv and it also confirmed the existence of the ghost for $1/3 < \lambda < 1$ but claimed that there is no strong
coupling problem. However, their argument is solely based on the fact that, in the $c_{\zeta} \to 0$ limit, the time kinetic term in the quadratic action diverges due to the factor $1/c_{\zeta}^2$. However, in order to see wether there is a strong coupling problem or not, one should look at interactions.

\begin{acknowledgments}
We thank Anzhong Wang and Takahiro Tanaka for useful discussions. KK would like to thank Sergei Sibiryakov for useful discussions that motivated us to study higher-order interactions of the scalar graviton. KK is supported by ERC, RCUK and STFC. FA is supported by a JSPS Research Fellowship.
\end{acknowledgments}

\appendix

\section{\label{App:lambdaone}$\lambda=1$ case}
In this Appendix, we consider the $\lambda=1$ case in the approach described in section III. In this case, we cannot determine $\beta$ from the momentum constraint. The quadratic action is obtained as
\begin{equation}
S= \int d^3 x dt \left(
-3 a^3 \dot{\zeta}^2 + a (\partial \zeta)^2 + 2 a \dot{\zeta} \partial^2 \beta
\right).
\end{equation}
As expected, now $\beta$ becomes a Lagrangian multiplier and we cannot determine $\beta $ in terms of $\zeta$. Then we get $\dot{\zeta}=0$ as a constraint. Let us apply the Hamiltonian formulation to see if this is consistent with the analysis in section II. The conjugate momentum is
\begin{equation}
\pi_{\zeta} = - 6 a^3 \dot{\zeta} + 2 a \partial^2 \beta, \quad \pi_{\beta}=0.
\end{equation}
The constraint $\dot{\zeta}=0$ implies
\begin{equation}
\partial^2 \beta = \frac{\pi_{\zeta}}{2 a}.
\end{equation}
Interestingly, this solution is the same as in the $c_{\zeta} \neq 0$ case.
The Lagrangian can be written as
\begin{equation}
{\cal L} = \pi_{\zeta} \dot{\zeta}- {\cal H}, \quad {\cal H} = -a (\partial \zeta)^2.
\end{equation}
This agrees with the result in section II. Using the constraint $\dot{\zeta}=0$, the cubic
interaction terms can be written as
\begin{equation}
S_3= \int dt d^3 x
\left[
a \zeta \partial_i \zeta \partial^i \zeta
+\frac{3}{2a} \zeta \Big(\partial_i \partial_j \beta\partial^i \partial^j \beta -(\partial^2 \beta)^2 \Big) - \frac{2}{a} \partial^2 \beta\partial_i \zeta\partial^i \beta
 \right].
\end{equation}
 So we find that the formulae (\ref{S2ham}) and (\ref{S3ham}) of section III can be applied even for the $c_{\zeta}=0$ case.

Let us study the solutions for perturbations about the Minkowski background. The Hamilton's equations
give $\dot{\zeta}=0$ and
\begin{equation}
\dot{\pi}_{\zeta} = -2 \partial^2 \zeta,
\end{equation}
which gives $\pi_{\zeta} = 2 f(x^i) - 2 t \partial^2 \zeta$, where $f(x^i)$ is an arbitrary
integration constant. This yields
\begin{equation}
\partial^2 \beta = f(x^i) - t \partial^2 \zeta(x^i).
\end{equation}
This agrees with the solution obtained in Ref.~\cite{Wang:2009yz}. As is discussed in \cite{Wang:2009yz},
the existence of the growing solution in time does not imply an instability as the dynamical variable
$\zeta$ is stable. The integration constant $f(x^i)$ is related to the ``enhanced symmetry'' discussed
in Ref.~\cite{Horava:2008ih}, that is, the theory is invariant under $\delta N_i = \partial \epsilon(x^i)$. However, this arises as a solution for the equation of motion and this does not change the dynamical degrees of freedom. Indeed, the dynamical variable $\zeta$ cannot be eliminated by this ``symmetry''. Thus even for $\lambda=1$, there still remains the scalar graviton.

\section{\label{App:Kinflation}Comparison to k-inflation models}
It is interesting to compare the actions for the scalar graviton to those for the curvature perturbations in k-inflation models with an arbitrary sound speed \cite{Seery:2005wm, Chen:2006nt}:
\begin{eqnarray}
S_2 &=& \int dt d^3 x
\left[a^3 \frac{\epsilon}{c_{s}^2} \dot{\zeta}^2 - a \epsilon (\partial \zeta)^2 \right],
\end{eqnarray}
\begin{eqnarray}
S_3 &=& \int dt d^3 x
\left[
-\epsilon a \zeta \partial_i \zeta \partial^i \zeta  +\frac{3 a^3 \epsilon}{c_{s}^2} \zeta \dot{\zeta}^2
+\frac{1}{2a}
 \left(3 \zeta - \frac{\dot{\zeta}}{H} \right)
 \Big(\partial_i \partial_j \beta\partial^i \partial^j \beta -(\partial^2 \beta)^2 \Big)
  \right. \nonumber\\
  && \left.
  - \frac{2}{a} \partial^2\beta\partial_i \zeta\partial^i \beta
-a^3 (\Sigma + 2 \tilde\lambda) \frac{\dot{\zeta}^3}{H^3}
 \right],
\end{eqnarray}
where
\begin{equation}
\beta =- \frac{\zeta}{H} + \xi, \quad \partial^2 \xi =a^2 \frac{\epsilon}{c_s^2} \dot{\zeta},
\end{equation}
 $\epsilon=-\dot{H}/H$ and $\Sigma$ and $\tilde \lambda$ are determined by the kinetic term of the scalar field whose detailed form is not important here. We find that these actions agree with our Eqs. (\ref{S2}) and (\ref{S3}) if we take $c_s \to 0$, $\epsilon =-1$ and $\dot{\zeta}/H =0$. The last condition is natural for a small sound speed as $\dot{\zeta} \propto c_s k \zeta$. This also implies that $\xi$ is dominant in the solution for $\beta$ and this again agrees with Eq.~(\ref{betasol}).
Note that in the k-inflation case, the action contains terms which have only one time derivative and we can perform a lot of integrations by parts to simplify the action. However, in our case, there is no term which has only one time derivative and we cannot reduce the action further.  The similarity between the two models is interesting and it deserve further investigations.


\end{document}